\documentclass[preprint,longbibliography,superscriptaddress,tightenlines,nofootinbib]{revtex4-1}

\newcommand{\be}{\begin{equation}}
\newcommand{\ee}{\end{equation}}
\newcommand{\bea}{\setlength\arraycolsep{2pt} \begin{eqnarray}}
\newcommand{\eea}{\end{eqnarray}}
\newcommand{\nn}{\nonumber}

\def\0{{\sst{(0)}}}
\def\1{{\sst{(1)}}}
\def\2{{\sst{(2)}}}
\def\3{{\sst{(3)}}}
\def\4{{\sst{(4)}}}
\def\5{{\sst{(5)}}}
\def\6{{\sst{(6)}}}
\def\7{{\sst{(7)}}}
\def\8{{\sst{(8)}}}
\def\sst#1{{\scriptscriptstyle #1}}

\usepackage{multirow}
\usepackage{xcolor}
\usepackage{bm}
\usepackage{ulem}
\usepackage{graphicx}
\usepackage{epsfig,psfrag}
\usepackage{amsmath}
\usepackage{amssymb}
\usepackage{amsbsy}
\usepackage{amsthm}
\usepackage{amsfonts}
\usepackage{wasysym}
\usepackage{bbm}
\usepackage{tabularx}
\usepackage{euscript}
\usepackage{enumerate}
\usepackage{amsfonts}
\usepackage{exscale}
\usepackage{bbold}
\usepackage{float}
\usepackage{slashed}
\usepackage{subfigure}
\usepackage{comment}
\usepackage[]{hyperref}

\numberwithin{equation}{section}

\def\bea{\begin{eqnarray}}
\def\eea{\end{eqnarray}}
\def\nn{\nonumber}
\def\ba{\begin{array}}
	\def\ea{\end{array}}
\def\nn{\nonumber}
\def\Tr{\text{Tr}}

\def\J{\mathcal{J}}

\def\L{\mathcal{L}}

\def\avg#1{\left\langle#1\right\rangle}
\def\bra#1{\left\langle#1\right|}
\def\ket#1{\left|#1\right\rangle}

\def\kc#1{\left(#1\right)}
\def\kd#1{\left[#1\right]}
\def\ke#1{\left\{#1\right\}}

\def\Im{{\rm Im}}

\def\be{\begin{equation}}       \def\ee{\end{equation}}
\def\bea{\begin{eqnarray}}      \def\eea{\end{eqnarray}}
\def\ba{\begin{array}}
	\def\ea{\end{array}}
\def\bnum{\begin{enumerate} }
	\def\enum{\end{enumerate}}

\def\nn{\nonumber}

\def\=>{\Rightarrow}
\def\>{\rightarrow}

\def\eye2{Fathbb{I}}

\def\Tr{\mathrm{Tr}}

\DeclareMathOperator\sech{sech}

\def\TT{{T\bar T}}
\def\Sch{{\text{Sch}}}

\begin{document}

\title{$T\bar T$ deformation on multiquantum mechanics and regenesis}

\author{Song He}
\email{hesong@jlu.edu.cn}
\affiliation{Center for Theoretical Physics, College of Physics, Jilin University,  Changchun 130012, People's Republic of China.}
\affiliation{Max Planck Institute for Gravitational Physics (Albert Einstein Institute), Am M\"uhlenberg 1, 14476 Golm, Germany.}

\author{Zhuo-Yu Xian}
\email{zhuo-yu.xian@physik.uni-wuerzburg.de}
\affiliation{Institute for Theoretical Physics and Astrophysics and W\"urzburg-Dresden Cluster of Excellence ct.qmat,\\
Julius-Maximilians-Universit\"at W\"urzburg, 97074 W\"urzburg, Germany}
\affiliation{Institute of Theoretical Physics, Chinese Academy of Science,\\ Beijing 100190, People's Republic of China.}

\begin{abstract}
We study the $T\bar T$ deformation on multiquantum mechanical systems. By introducing dynamical coordinate transformation, we reformulate the one-dimensional $T\bar T$ deformation of generic quantum mechanical systems, which is consistent with the previous proposal in the literature. We further study the thermo-field-double state under the $T\bar T$ deformation on these systems, which include the conformal quantum mechanical system, the Sachdev-Ye-Kitaev model, and the model satisfying eigenstate thermalization hypothesis. We find common regenesis phenomena in which the signal injected into one local system can regenerate from the other local system. From the $AdS_2/CFT_1$ perspective, we study the deformation of Jackiw-Teitelboim gravity governed by Schwarzian action and find that these regenesis phenomena are realized by exchanging boundaries graviton via the nonlocal $T\bar T$ coupling.

\end{abstract}

\maketitle

\tableofcontents

\section{Introduction}

The $\TT$ deformation of field theory has recently attracted significant research interest in field theory and holographic duality. The $\TT$ deformation of two-dimensional (2D) rotational and translational invariant field theory have been defined in previous research \cite{Zamolodchikov:2004ce,Smirnov:2016lqw,Cavaglia:2016oda},
as being triggered by the irrelevant and double-trace operator $T\bar{T}=-\operatorname{det}\left(T_{\mu \nu}\right)$. Although the $\TT$ deformation flows toward ultraviolet (UV), it exhibits numerous intriguing properties, particularly its integrability \cite{Smirnov:2016lqw,LeFloch:2019wlf,Jorjadze:2020ili}. If the undeformed theory is integrable, a set of infinite commuting conserved charges or Korteweg-de-Vries charges exists in the deformed theory. If the theory is maximally chaotic, the deformed theory holds the maximal chaos \cite{He:2019vzf, He:2020qcs}, which agrees with the $\TT$ deformation is irrelevant.

The $\TT$ deformation of the (0+1)-dimensional quantum mechanical (QM) system is studied in \cite{Gross:2019ach,Gross:2019uxi}. When the QM system is taken as the Sachdev-Ye-Kitaev (SYK) model \cite{Sachdev:1992fk,Kitaev:2015smq,Kitaev:2015hch,Sachdev:2015efa,Maldacena:2016hyu,Kitaev:2017awl}, the deformed SYK model exhibits the maximal chaotic behavior as the undeformed model. Moreover, one-dimensional deformation of boson gas has been studied in \cite{Jiang:2020nnb}.

The present study analyzes the $\TT$ deformation in multi-QM systems.
It is a broad class of QM's integrable deformations, which can be regarded as a transformation of the Hamiltonian $H\to f(H)$.
As shown in \cite{Cavaglia:2016oda}, the $\TT$ deformation on multi-QM systems effectively couple the local system and generate a nonloca phenomenon. In this paper, we calculate the causal correlation caused by the $\TT$ deformation on the bi-QM system, in which the two local QM systems, labeled $L$ and $R$, share the Hamiltonian in the same form.

We will focus on a particular entangled state in the bi-QM system, the thermo-field-double (TFD) state, in which the local system is in a thermal state, and the local entropy is caused by entanglement. When the QM system has holographic duality, the geometric correspondence of the TFD state is an eternal black hole \cite{Maldacena:2001kr, Maldacena:2013xja}. 

When the two QM systems are coupled with each other, and their interactions match the entanglement structure of the TFD state, a phenomenon similar to quantum teleportation appears, in which the signal injected into one QM system can regenerate from the other QM system \cite{Gao:2018yzk}. The teleportation of the quantum state is constructed in the SYK model \cite{Gao:2019nyj}, and in 2D conformal field theory (CFT) \cite{Numasawa:2016emc}. We call this phenomenon regenesis.

The geometric correspondence is known as a traversable wormhole \cite{Gao:2016bin, Maldacena:2017axo, Maldacena:2018lmt, Garcia-Garcia:2019poj}, in which a signal injected into the external black hole from one boundary at a proper time can transverse the Einstein-Rosen bridge and reach the other boundary. The traversability of the wormhole is closely associated with the violation of averaged null energy condition (ANEC). The ANEC indicates that the integral of null energy on null ray must be non-negative in any UV complete quantum field theory. The ANEC has been proven in many particular cases \cite{Kelly:2014mra, Faulkner:2016mzt, Hartman:2016lgu}.
The ANE can measure changes in the causal structure when the matter stress tensor perturbs the solution of the vacuum Einstein equation. When the ANE is negative, the null ray in the unperturbed metric becomes timelike in the perturbed metric. In classical general relativity, the existence of a traversable wormhole implies negative ANE. To construct a traversable wormhole, the authors of \cite{Gao:2016bin, Maldacena:2017axo} added the double-trace deformations $O_LO_R$ between the two sides of the black hole. Under this deformation, the ANEC is violated, and the Einstein-Rosen bridge of the eternal black hole becomes traversable.

However, not all the regenesis phenomena have geometric correspondence in semiclassical approximation \cite{Maldacena:2017axo, Gao:2018yzk}. The signal is injected earlier than the scrambling time in the interference region. The backreaction to the wormhole destroys the correlation between $O_L$ and $O_R$ and contributes a nonzero phase to the correlator carrying the signal. The signal regenerates from the other side at the time-reversed to the injection time. Such a regenesis phenomenon is called a ``quantum traversable wormhole'' \cite{Gao:2018yzk}.

The above double-trace deformation on the bi-QM system is relevant and can change the ground state \cite{Maldacena:2018lmt, Garcia-Garcia:2019poj}. The present study considers the $\TT$ deformation of bi-QM systems in the TFD state. As a double-trace deformation with stress tensors, it is nonlocal and irrelevant. So we expect to find regenesis phenomena contributed by UV channels. In the usual construction of a traversable wormhole, the nonlocal deformation should match the entanglement structure of the TFD state such that the $O_L$ and $O_R$ constructing the deformation should be initially correlated. However, the $\TT$ deformation is unique and unrelated to the entanglement structure; therefore, we expect a relatively weak but general regenesis phenomenon.

The organization of this paper is as follows. In Sec.~\ref{Sec:GeneralTTbar}, we give a general framework of the $\TT$ deformation of single or multi-QM systems. In Sec.~\ref{Sec:QM}, we study the first-order $\TT$ deformation of bi-QM systems in TFD states. Taking conformal QM, the SYK model, and the system satisfying the eigenstate thermalization hypothesis (ETH), we discuss general regenesis phenomena in which a signal can pass from one QM system to another QM system. In Sec.~\ref{Sec:Schwarzian}, we study the $\TT$ deformation in a wormhole based on Schwarzian theory, the results of which agree with those of the bi-QM system analysis. A summary and a discussion of prospects are given in Sec.~\ref{Sec:Summary}, which concludes the paper.

\section{$\TT$ deformation on $(0+1)$-dimensional systems}\label{Sec:GeneralTTbar}

In this section, we give some general approaches to study the $\TT$ deformation on a multi-QM system.

\subsection{Solution of $\TT$ deformed Hamiltonian}

Consider a pair of canonical variables $\ke{q,p}$ and a Hamiltonian $H_{0}(q,p)$. Given a solution of the Hamiltonian equation
\begin{align}
q(t)=\tilde{q}\left(q_{0}, p_{0}, t\right) ,\quad p(t)=\tilde{p}\left(q_{0}, p_{0}, t\right),
\end{align}
in which the initial conditions are $q_{0}=q(0)$ and $p_{0}=p(0)$, we consider a new Hamiltonian
\begin{align}
H=f\left(H_{0}\right)
\end{align}
that may be in the form of the $\TT$ deformed Hamiltonian proposed in Refs.~\cite{Gross:2019ach,Gross:2019uxi}. The new Hamiltonian equations are
\begin{align}
q'=f'\left(H_{0}\right) \frac{\partial H_{0}}{\partial p}, \quad p'=-f'\left(H_{0}\right) \frac{\partial H_{0}}{\partial x}.
\end{align}
The solution of the deformed theory with the same initial condition is
\begin{align}
q(t)=\tilde{q}\left(q_{0}, p_{0},T\right), \quad
p(t)=\tilde{p}\left(q_{0}, p_{0}, T\right),\quad
T= f'\left(H_{0}\left(q_{0}, p_{0}\right)\right) t,
\end{align}
where $T$ is the dynamical coordinate.

For the theory $H_0(\vec q,\vec p)$ with multiple pairs of canonical variables $\{\vec q=(q_1,q_2,...,q_n)$, $\vec p=(p_1,p_2,...,p_n)\}$, similarly, we can construct a new solution
\begin{align}\label{DeformedSolution}
q_s(t)=\tilde{q_s}\left(\vec q_{0}, \vec p_{0},T\right), \quad
p_s(t)=\tilde{p_s}\left(\vec q_{0}, \vec p_{0}, T\right),\quad
T= f'\left(H_{0}\left(\vec q_{0}, \vec p_{0}\right)\right) t,\quad s=1,2,...,n,
\end{align}
which satisfies the initial conditions $\vec q_0=(q_1(0),q_2(0),...,q_n(0))$ and $\vec p_0=(p_1(0),p_2(0),...,p_n(0))$.

\subsection{$\TT$ deformation and dynamical coordinate}

In this section, we realize the $\TT$ deformation in the $(0+1)$ dimension by generalizing the dynamical coordinate transformation from Refs.~\cite{Dubovsky:2018bmo,Dubovsky:2017cnj}. One can also refer to recent extensive studies~\cite{Cardy:2018sdv,Conti:2018tca,Aguilera-Damia:2019tpe,Tolley:2019nmm,Mazenc:2019cfg,Ouyang:2020rpq,Caputa:2020lpa} in the $(0+1)$ dimension. We couple the original action $S_{0}$ to a $(0+1)$-dimensional ``gravity'' as
\begin{align}
S\left[e_{\mu}, v^{\mu}, \phi\right] &=S_{\text {grav }}\left[e_{\mu}, v^{\mu}\right]+S_{0}\left[e_{\mu}, \phi\right] \\
S_{\text {grav }}\left[e_{\mu}, v^{\mu}\right] &=\frac{1}{\lambda} \int d t e_{t} B\left(e_{t} v^{t}\right),
\end{align}
with $e_{\mu}$ the dynamical tetrad, $B$ the undetermined function, and $v^{\mu}$ a fixed co-tetrad corresponding to the metric on which the deformed theory lives. We can take $v^{t}=1$ and then have
\begin{align}\label{TCoordinate}
v^{T}=\frac{d T}{d t}, \quad e_{T}=1, \quad e_{t}=\frac{d T}{d t}.
\end{align}
Take the scalar theory
\begin{equation}\label{S0}
S_{0}=\int d t e_{t}\left(\frac{1}{2\left(e_{t}\right)^{2}} \partial_{t} \phi \partial_{t} \phi-V(\phi)\right),
\end{equation}
as an example. By introducing the canonical momentum $p$, we can write it into a first-order form
\begin{align}
S_{0}=\int d t e_{t}\left(\frac{1}{e_{t}} p \partial_{t} \phi-H_{0}(\phi, p)\right),
\end{align}
with $H_{0}$ the undeformed Hamiltonian.

The equation of motion of $e_{t}$ is
\begin{equation}\label{EoMTetrad}
e_tv^{t} B^{\prime}\left(e_{t} v^{t}\right)+B\left(e_{t} v^{t}\right)-\lambda H_{0}=0.
\end{equation}
In the $T$ coordinate, from (\ref{TCoordinate}), it becomes
\begin{align}
\frac{d T}{d t} B^{\prime}\left(\frac{d T}{d t}\right)+B\left(\frac{d T}{d t}\right)-\lambda H_{0}
=f'(H_0)B^{\prime}\left(f'\left(H_{0}\right)\right)+B\left(f'\left(H_{0}\right)\right)-\lambda H_{0}
=0,
\end{align}
where $d T=f'\left(H_{0}\right) d t$ is used. 
It could be solved by
\begin{align}\label{Bf}
B\left(f'(H)\right)=\lambda H-\frac{\lambda f(H)}{f'(H)}+\frac{C}{f'\left(H\right)},
\end{align}
where $C$ is a constant of integration.

In the $t$ coordinate, one can find the solution $B$ of (\ref{EoMTetrad}) such that $e_{t}=f'\left(H_{0}\right)$.  By integrating out $e_{t}$ in the action, the resulting action is 
\begin{align}
S=\int d t\left(p \partial_{t} \phi-f\left(H_{0}\right)\right),
\end{align}
where the constant term $C/\lambda$ has been dropped.

For $T \bar{T}$ deformation \cite{Gross:2019uxi}, we have
\begin{align}\label{TTbar}
f(H)=\frac{1-\sqrt{1-8 H \lambda}}{4 \lambda}.
\end{align}
The function $B$ is determined as
\begin{align}
B(x)=\frac{(x-1)^{2}}{8 x^{2}}.
\end{align}
Finally, one can check that the deformed Hamiltonian satisfies the flow equation
\be \label{floweq}
2\partial_\lambda H=\frac{H^2}{4-2\lambda H},
\ee
which is consistent with \cite{Gross:2019uxi}.
If $S_{0}$ takes the form given by (\ref{S0}) , the deformed action after integrating out $e_{t}$ is given by
\begin{align}
S=\int d t\left(\frac{\sqrt{4 \partial_{t} \phi \partial_{t} \phi \lambda+1} \sqrt{1-8 \lambda V(\phi)}-1}{4 \lambda}\right).
\end{align}
We can apply the above approach to the single 1D Liouville action and obtain the deformed action given in \cite{Gross:2019uxi}.

We follow the dynamical coordinate transformation proposed by \cite{Dubovsky:2018bmo,Dubovsky:2017cnj} in 2D quantum field theories to realize the $T\bar{T}$ flow equation. We extend this approach to 1D, namely, the undeformed theory couples with the 1D massive gravity $B$, which can be regarded as an alternative way to realize the $T\bar{T}$ deformation. We introduce an undetermined function $B$ to characterize the unclear massive gravity in 1D. Our approach gives the same results as Ref.~\cite{Gross:2019ach}. But we work in the Minkowski signature and use the saddle point approximation, while the authors in Ref.~\cite{Gross:2019ach} work in the Euclidean signature and use the exact path integral.

\subsection{$\TT$ deformation on multifields}\label{Sec:Liouville}

For the theory with multiscalars $\vec\phi=(\phi_1,\phi_2,...,\phi_n)$ in the (0+1) dimension, the $\TT$ deformed action can be obtained as follows. We consider the original Lagrangian
\begin{align}
\mathcal L_0=  \frac12 \sum_s\phi_s'\phi_s'- V(\vec\phi).
\end{align}
Then the Hamiltonian is
\begin{align}
H_0=\frac12\sum_s p_sp_s+V(\vec\phi),
\end{align}
where $p_s$ is the momentum conjugate to $\phi_s$.
We consider the $\TT$ deformation as
\begin{align}
H_\lambda=\frac{1-\sqrt{1-8\lambda H_0}}{4\lambda}.
\end{align}
Then the deformed Lagrangian is
\begin{align}\label{TTLag}
\mathcal L_\lambda=\frac{\sqrt{(1+4\lambda \sum_s\phi_s'\phi_s')(1-8\lambda V(\vec\phi))}}{4\lambda}.
\end{align}
It satisfies the flow equation
\begin{align}
\frac{\partial \mathcal L_\lambda}{\partial \lambda}=\frac{- T_\lambda^2}{1/2-2\lambda T_\lambda},
\end{align}
where the deformed energy-momentum tensor is
\begin{align}  
T_\lambda=\sum_s\phi_s' \frac{\partial\mathcal L_\lambda}{\partial \phi_s'}-\mathcal L_\lambda.
\end{align}

\section{Causal correlation caused by the $\TT$ deformation}\label{Sec:QM}

\subsection{First-order $\TT$ deformation on bi-QM system}

We consider a QM system with Hilbert space $\mathcal H$ and Hamiltonian $H$. Denote the dimension of the Hilbert space as $D=\dim\mathcal H$ and the spectrum density of $H$ as $\rho(E)$. The summation of the energy spectrum can be written as an integral forms
$
\sum_E=D\int dE \rho(E).
$

Now, we consider the two copies of the QM system and call them QM$_L$ and QM$_R$. The Hilbert space is $\mathcal H\otimes\mathcal H$. The Hamiltonian is $H_0=H_L+H_R$, where $H_L=H\otimes 1$ and $H_R=1\otimes H$.

We consider the global Hamiltonian as
\begin{align}
H_\lambda=f(H_0),
\end{align}
with
\begin{align}
f(H)=H+2\lambda H^2,
\end{align}
which is the $\TT$ deformation (\ref{TTbar}) at the first order.
The $\TT$ term couples QM$_L$ and QM$_R$ nonlocally. Because the $\TT$ deformation is irrelevant, new mechanics is introduced in the UV, but the ground state of the deformed theory generally remains unchanged. We introduce the $\TT$ deformation of states with two strategies.

\subsection{$\TT$ quenched TFD state}

\subsubsection{State}

We prepare the undeformed and non-normalized TFD state as
\begin{align}
\ket{ \Psi}=\sum_E e^{-\beta E/2}\ket{E}_L\ket{E}_R
\end{align}
in which the reduced density matrix on each side is
\begin{align}
\rho= \sum_E  e^{-\beta E}\ket{E}\bra{E}.
\end{align}
Their normalization is
\begin{align}
\ket{\tilde\Psi}=\ket{\Psi}/\sqrt{Z(\beta)},\quad
\tilde\rho=\rho/Z(\beta),\quad
Z(\beta)=\sum_E e^{-\beta E}.
\end{align}
We consider the TFD state $\ket{\Psi}$ at $t=0$ and let it evolve with the deformed Hamiltonian $H_\lambda$, namely,
\begin{align}\label{QuenchedTFD}
\ket{\Psi(t)}
=e^{-i t f(H_0)}\ket{\Psi}.
\end{align}
Notably, the reduced density matrix on each side remains unchanged, namely,
\begin{align}\label{QuenchedLocal}
\rho(t)=\rho.
\end{align}
Therefore, the entanglement between QM$_L$ and QM$_R$ is independent of the time.

\subsubsection{Correlation}\label{Sec:Correlation}

Consider a local and Hermitian operator $O$ acting on $\mathcal H$. Its two copies are
\begin{align}
O_L=O\otimes1,\quad O_R=1\otimes O^T,
\end{align}
where the transpose is taken on the energy basis of $H_0$.

To study the causal correlation between two QM systems under the $\TT$ quench, we calculate the retarded correlator
\begin{align}
G^R_{LR}(t_1,t_2)=-i\Theta(t_-)\bra{\tilde\Psi}[O_L(t_1),O_R(t_2)]\ket{\tilde\Psi}=2\Theta(t_-)\Im \bra{\tilde\Psi}O_L(t_1) O_R(t_2)\ket{\tilde\Psi},
\end{align}
where $ t_\pm=t_1\pm t_2$ and $O(t)= e^{it H_\lambda}Oe^{-it H_\lambda}$.
This is the linear response of the protocol, which is sending a signal from QM$_R$ at time $t_2$ and measuring QM$_L$ at time $t_1$. We consider $t_-\geq0$ below. We first calculate the correlator on the energy basis as
\begin{align}
&\bra{\Psi}O_L(t_1) O_R(t_2)\ket{ \Psi}\\
=&\sum_{E_1E_2}O_{12}O_{21}\exp\ke{-\frac\beta2 E_1-\frac\beta2 E_2+it_1f(2E_1)-it_2f(2E_2)-it_{12}f(E_1+E_2)}\\
=&\sum_{E_1E_2}O_{12}O_{21}\exp\ke{-\frac\beta2 E_+ + 2i\lambda t_-E_-^2 + it_+ (1+4\lambda E_+)E_- }\\
=&\sum_{E_1E_2}O_{12}O_{21}\exp\ke{-\frac\beta2 E_+} \int_{-i\infty}^{i\infty} d \beta' K(-2i\lambda t_-,it_+(1+4\lambda E_+)+\beta')\exp\ke{-\beta' E_-},
\end{align}
where $O_{ij}=\bra{E_i}O\ket{E_j},\ E_\pm=E_1\pm E_2$, and the kernel
\begin{align}
K(\alpha,\beta)
=\frac1{2\pi i}\int_{-\infty}^{\infty} dE e^{-\alpha E^2+\beta E}
=\frac{-i}{2 \sqrt{\pi\alpha }} \exp {\frac{\beta ^2}{4 \alpha }}.
\end{align}

To analytically calculate the above transformation, we consider the weakly coupled limit $|\lambda|\ll 1/E_\beta$, where $E_\beta=\bra{\tilde\Psi}H_s\ket{\tilde\Psi}$ is the energy at the inverse temperature $\beta$. Thus, we can approximate $|\lambda| E_+\ll 1$ such that
\begin{align}\label{DeformedG}
G_{LR}(t_1,t_2)
=\bra{\tilde\Psi}O_L(t_1) O_R(t_2)\ket{\tilde\Psi}
\approx i \int_{-\infty}^{\infty} du K(-2i\lambda t_-,it_++iu) G_W(u;\beta),
\end{align}
where the half-circle Wightman correlator is $G_W(u;\beta)=\Tr[e^{-(\beta/2+iu)H}Oe^{-(\beta/2-iu)H}O]/Z(\beta)$. The approximation becomes exact when $t_+=0$ or $t_-\to\infty$.
The complex conjugate of (\ref{DeformedG}) shows
\begin{align}
\lambda\leftrightarrow-\lambda,\quad G_{LR}(t_1,t_2)\leftrightarrow G_{LR}(t_1,t_2)^*
\end{align}
at the weakly coupled limit. Thus, we consider a positive $\lambda$ value.

Furthermore, at weakly coupled limit $|\lambda|\ll 1/E_\beta$, we can use the saddle point approximation $u=-t_++\delta u$ in (\ref{DeformedG}) when the variance $\sqrt{\lambda t_-}$ in the exponent is small compared to the characteristic time in $G_W$, namely,
\begin{align}
G_{LR}(t_1,t_2)
\approx&  \int_{-\infty}^{\infty} d\delta u \sqrt{\frac{i}{8\pi \lambda  t_- }} \exp {\frac{-i\delta u ^2}{8 \lambda t_- }} \kc{G_W(-t_+;\beta)+\frac12 \delta u^2 G_W''(-t_+;\beta)}\\
=&G_W(-t_+;\beta) - 2i \lambda t_- G_W''(-t_+;\beta).
\end{align}
Thus, the retarded correlator is approximated by
\begin{align}\label{GLRSaddle}
G^R_{LR}(t_1,t_2)\approx  -4 \lambda t_- \Theta(t_-) G_W''(-t_+;\beta).
\end{align}
If the characteristic time in $G_W$ is $\beta$, such as conformal correlators, the valid region of the approximation is $\lambda t_-\ll\beta^2$.

This is also the result from the first-order perturbation on $\lambda$, since
\begin{align}\label{FirstOrderLambda}
[O_L(t_1),O_R(t_2)]
=-4i\lambda t_- \dot O_L^{(0)}(t_1)\dot O_R^{(0)}(t_2)+\mathcal O[\lambda^2],
\end{align}
where $O^{(0)}(t)= e^{it H_0}O e^{-it H_0}$. Because of the entanglement structure, $G''_W(-t_+;\beta)$ is maximized at $t_+=0$. Therefore, the signal appears from QM$_L$ near the time $t_1=-t_2$.

A similar regenesis phenomenon appears if we apply an instantaneous $\TT$ quench on the TFD state
\begin{align}
H_\lambda(t)=H_L+H_R+2\lambda (H_L+H_R)^2 \delta(t).
\end{align}
The retarded correlator at the first-order perturbation of $\lambda$ is
\begin{align}
G^R_{LR}(t_1,t_2)
=&-i\Theta(t_-)\bra{\tilde\Psi}\kd{e^{i2\lambda(H_L+H_R)^2}O^{(0)}_L(t_1)e^{-i2\lambda(H_L+H_R)^2},O^{(0)}_R(t_2)}\ket{\tilde\Psi} \label{GLRInstant}\\
\approx& -4\lambda \Theta(t_-) \bra{\tilde\Psi} \dot O_L^{(0)}(t_1)\dot O_R^{(0)}(t_2) \ket{\tilde\Psi} \label{GLRInstantPerturb}\\
=& -4\lambda \Theta(t_-) G''_W(t_+;\beta). \label{GLRInstantResult}
\end{align}
When $t_1=-t_2=t>0$, $G^R_{LR}(t,-t)\approx -4\lambda G''(0;\beta)$ is completely independent of $t$. The signal can instantly pass through the system from QM$_R$ to QM$_L$ .

Both kinds of $\TT$ quench lead to a nonvanishing retarded correlator. The entanglement structure of the TFD state leads to the quantum correlation between the operator $O_L$ and $O_R$. Because the operators also perturb the energy correlation, under $\TT$ deformation, the quantum correlation becomes the causal correlation. It can be described as sending a signal into QM$_R$ at a particular time and measuring it on QM$_L$ at the reverse time with the highest intensity, similar to the traversal phenomenon under nonlocal double-trace deformation in the interference region discussed in \cite{Gao:2018yzk, Gao:2019nyj}. However, there is some difference between ours and theirs. The double-trace deformations in their setting are usually relevant, which changes IR physics. At the same time, the $\TT$ deformation is irrelevant, which only changes the UV physics. So our regenesis could happen instantly and without a finite waiting time. More specifically, since the $G^R_{LR}$ is related to the two-point function $G_W$ rather than four point functions, namely out-of-time-order correlator, it does not rely on chaos and is not associated with the scrambling \cite{Sekino:2008he, Shenker:2013pqa, Kitaev:2015hch}.

\subsection{$\TT$ deformed TFD state}

\subsubsection{State}

Alternatively, we can prepare a new TFD state with the $\TT$ deformed Hamiltonian
\begin{align}
\ket{\Psi_\lambda}=\sum_E e^{-\beta f(E)/2}\ket{E}_L\ket{E}_R,\\
\rho_{\lambda}= \sum_E  e^{-\beta f(E)}\ket{E}\bra{E}.
\end{align}
It evolves with the deformed Hamiltonian as
\begin{align}\label{DeformedTFD}
\ket{\Psi_\lambda(t)}
=e^{-i t f(H_0)}\ket{\Psi_\lambda},\\
\rho_\lambda(t)=\rho_\lambda, \label{DeformedLocal}
\end{align}
where $\rho_\lambda$ is the reduced density matrix on each side. The state can be normalized as $\ket{\tilde\Psi_\lambda}=\ket{\Psi_\lambda}/\sqrt{Z_\lambda(\beta)}$, where the deformed partition function is $Z_\lambda(\beta)=\Tr[e^{-\beta f(H)}]$. The entanglement is time independent as well. Since $f'(E)>1$ for $\lambda>0$, the deformation enhances the imbalance of the energy distribution $e^{-\beta f(E)}$ such that low energy states have higher probabilities. Therefore, at the same temperature, the entanglement in the $\TT$ deformed TFD state is generally lower than that in the $\TT$ quenched TFD state.

\subsubsection{Correlation}
The correlator on the $\TT$ deformed TFD state is
\begin{align}
&\bra{\Psi_\lambda}O_L(t_1) O_R(t_2)\ket{ \Psi_\lambda}\\
=&\sum_{E_1E_2}O_{12}O_{21}\exp\ke{-\frac\beta2 f(E_1)-\frac\beta2 f(E_2)+it_1f(2E_1)-it_2f(2E_2)-it_{12}f(E_1+E_2)}\\
=&\sum_{E_1E_2}O_{12}O_{21}\exp\ke{-\frac\beta2 E_+(1+\lambda E_+) + 2\lambda \kc{it_--\frac\beta4}E_-^2 +  it_+ (1+4\lambda E_+)E_-}\\
=&\sum_{E_1E_2}O_{12}O_{21}\exp\ke{-\frac\beta2 E_+(1+\lambda E_+)}\nn \\
&~~~~~\int_{-i\infty}^{i\infty} d \beta' K\kc{-2\lambda\kc{it_--\frac\beta4},it_+(1+4\lambda E_+)+\beta'}\exp\ke{-\beta' E_-}.
\end{align}
Similarly, at the weakly coupled limit $|\lambda|\ll 1/E_\beta=1/\bra{\tilde\Psi_\lambda}H_s\ket{\tilde\Psi_\lambda}$, we use the approximation $|\lambda| E_+\ll 1$ and find
\begin{align}
G_{LR}(t_1,t_2)_\lambda
=\bra{\tilde\Psi_\lambda}O_L(t_1) O_R(t_2)\ket{\tilde\Psi_\lambda}
\approx i \int_{-\infty}^{\infty} du K\kc{-2\lambda\kc{i t_--\frac\beta4},it_++iu} G_W(u;\beta),
\end{align}
which is coincident with the $G_{LR}(t_1,t_2)$ in (\ref{DeformedG}) with the replacement $it_-\to it_--\frac\beta4$. From (\ref{FirstOrderLambda}), at the first-order perturbation on $\lambda$, the retarded correlator $G^R_{LR}(t_1,t_2)_\lambda$ is the same as $G^R_{LR}(t_1,t_2)$.

\subsection{Applications}

\subsubsection{Conformal QM}

\begin{figure}
	\centering
	\includegraphics[height=0.25\linewidth]{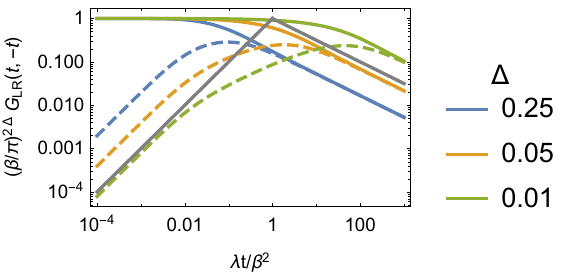}
	\caption{$G_{LR}(t,-t)$ for conformal QM, in which the real (imaginary) part is denoted as solid (dashed) line. The gray lines denote power laws $t$ and $t^{-1/2}$.}
	\label{fig:GLRCA}
\end{figure}

\begin{figure}
	\centering
	\includegraphics[height=0.25\linewidth]{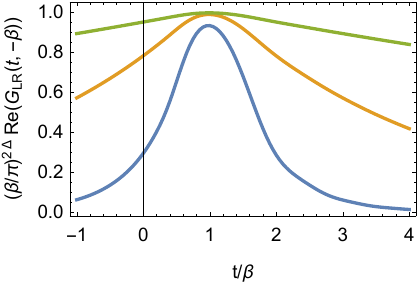}~~
	\includegraphics[height=0.25\linewidth]{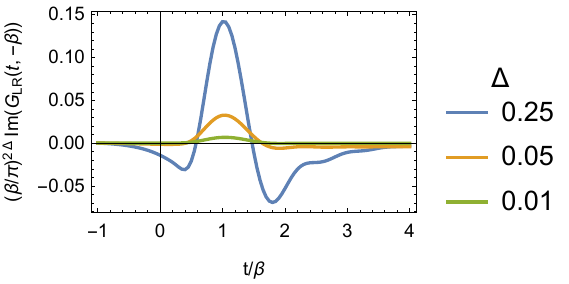}
	\caption{$G_{LR}(t,-\beta)$ for conformal QM, where $\lambda/\beta=0.01$.}
	\label{fig:GLRCDs}
\end{figure}

\begin{figure}[!htb]
	\centering
	\includegraphics[height=0.25\linewidth]{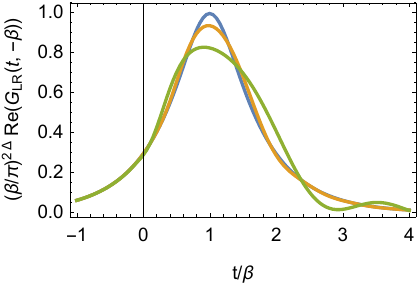}~~
	\includegraphics[height=0.25\linewidth]{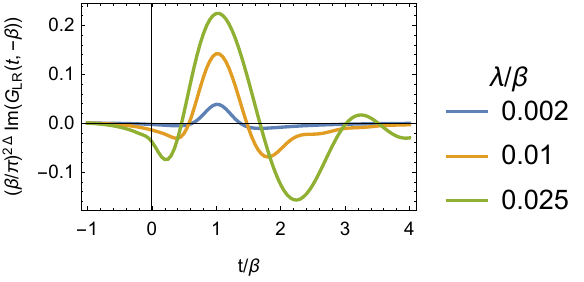}
	\caption{$G_{LR}(t,-\beta)$ for conformal QM, where $\Delta=0.25$.}
	\label{fig:GLRCls}
\end{figure}

We can apply the above formula to a conformal QM. For a primary operator $O$ with dimension $\Delta$, the Wightman correlator is
\begin{align}
G_W(t;\beta)=\kc{\frac{\pi}{\beta}\sech\frac{\pi t}{\beta }}^{2 \Delta }.
\end{align}
From (\ref{DeformedG}), the correlator on the $\TT$ quenched TFD state is
\begin{align}
G_{LR}(t_1,t_2)
=\kc{\frac{\pi}{\beta}}^{2\Delta}\sqrt{\frac i {8\pi x}}\int_{-\infty}^{\infty} du  \sech^{2\Delta}\kc{\pi u} \exp\frac{(u+t_+/\beta)^2}{i8x},\quad
x=\frac\lambda\beta\frac{t_-}{\beta},
\end{align}
as shown in Figs.~\ref{fig:GLRCA}, \ref{fig:GLRCDs} and \ref{fig:GLRCls}. In Fig.~\ref{fig:GLRCA}, the peak appears near the timescale $\beta^2/\lambda$, which indicates the best regenesis.
The correlator on the $\TT$ deformed TFD state is
\begin{align}
G_{LR}(t_1,t_2)_\lambda
=\kc{\frac{\pi}{\beta}}^{2\Delta}\sqrt{\frac i {8\pi x}}\int_{-\infty}^{\infty} du  \sech^{2\Delta}\kc{\pi u} \exp\frac{(u+t_+/\beta)^2}{i8x},\quad
x=\frac\lambda\beta \kc{\frac{t_-}{\beta}+\frac i4}.
\end{align}
{The behavior is close to that in the case of the $\TT$ quenched TFD state, except that the correlation is slightly suppressed due to the loss of entanglement.}

\subsubsection{The SYK model}

\begin{figure}
	\centering
	\includegraphics[height=0.25\linewidth]{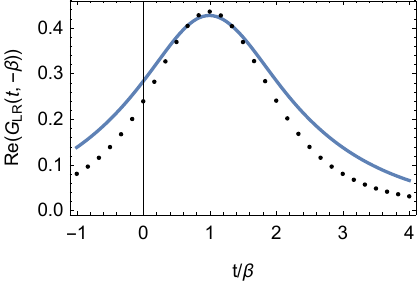}~~
	\includegraphics[height=0.25\linewidth]{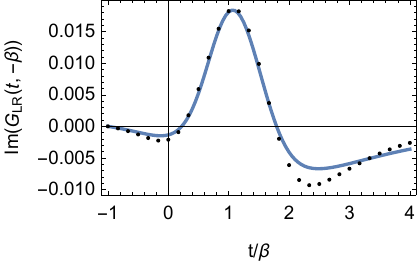}
	\caption{$G_{LR}(t,-\beta)$ for the SYK model. The dots denote the results from exact diagonalization. The curves denote the results from (\ref{QuenchedGSYK}). The parameters are $q=4,N=20,\J=1,\beta=2,\lambda=0.02$.}
	\label{fig:GLRCSYK}
\end{figure}

We consider the SYK model as the QM system, in which the local Hamiltonian is \cite{Sachdev:2015efa,Kitaev:2015hch,Maldacena:2016hyu,Kitaev:2017awl}
\begin{align}
H = \frac{i^{\frac{q}2}}{q!} \sum_{j_1,\cdots, j_q} J_{j_1,\cdots,j_q} \psi^{j_1}\cdots\psi^{j_q}, \quad \overline{J_{j_1,\cdots,j_q}^2} = \frac{2^{q-1}  (q-1)! \J^2}{q N^{q-1}}.
\end{align}
The SYK model exhibits free fermionic behavior in the UV and conformal symmetry in the infrared (IR). The two-point function interpolating between UV and IR can be solved at the large-$q$ limit. The Wightman function is \cite{Maldacena:2016hyu}
\begin{align}
G_W(t;\beta)=\frac12\kd{\kc{\frac{\cos\frac{\pi v}{2}}{\cosh\frac{\pi v t}{\beta}}}^2}^{1/q},\quad \pi v=\beta \J \cos\frac{\pi v}{2}.
\end{align}

For the $\TT$ quenched TFD state, the correlator at weakly coupled limit $|\lambda|\ll 1/E_\beta\sim \beta \J^2/N$ is similar to the conformal result:
\begin{align}\label{QuenchedGSYK}
G_{LR}(t_1,t_2)
= \kc{\frac{\pi v}{\beta\J}}^{2/q}\frac12\sqrt{\frac i {8\pi x}}\int_{-\infty}^{\infty} du  \sech^{2/q}\kc{\pi v u} \exp\frac{(u+t_+/\beta)^2}{i8x},\quad
x=\frac\lambda\beta\frac{t_-}{\beta},
\end{align}
which is close to the result from exact diagonalization in Fig.~\ref{fig:GLRCSYK}.
For the $\TT$ deformed TFD state, the correlator is
\begin{align}\label{DeformedGSYK}
G_{LR}(t_1,t_2)_\lambda
= \kc{\frac{\pi v}{\beta\J}}^{2/q}\frac12\sqrt{\frac i {8\pi x}}\int_{-\infty}^{\infty} du  \sech^{2/q}\kc{\pi v u} \exp\frac{(u+t_+/\beta)^2}{i8x},\quad
x=\frac\lambda\beta\kc{\frac{t_-}{\beta}+\frac i4}.
\end{align}

\subsubsection{The system satisfying the ETH}

We can apply the above formula to the system satisfying the ETH. Consider the Hermitian operator $O$ which satisfies \cite{Feingold:1986fu,Deutsch:1991qsm,Srednicki:1994cqt,Srednicki:1995qcs}  
\begin{align}
O_{ab}\approx\frac1{\sqrt D} F(E_+,E_-)R_{ab},\quad a\neq b,\\
\avg{R_{ab}}=0,\quad \avg{R_{ab}R_{cd}^*}=\delta_{ac}\delta_{bd},\quad E_\pm=E_a\pm E_b,
\end{align}
where $R_{ab}$ is a random matrix. We further assume that the operator $O$ has bandwidth $\Gamma$:
\begin{align}
F(E_+,E_-)\sim A(E_+/2) e^{-|E_-|/\Gamma}.
\end{align}
The off-diagonal part of the correlator of the operator on the $\TT$ quenched TFD state is
\begin{align}
&\bra{\Psi}O_L(t_1) O_R(t_2)\ket{ \Psi}-\sum_{a}O_{aa}^2\exp\ke{-\beta E_a}\\
=&\frac1D\sum_{a\neq b}\avg{R_{ab}R_{ba}}A\kc{\frac{E_+}{2}}^2\exp\ke{-\frac\beta2 E_+ + 2i\lambda t_- E_-^2 +  it_+ (1+4\lambda E_+)E_- -\frac2\Gamma |E_-|}\\
\approx&\frac1D\sum_{a\neq b}A\kc{\frac{E_+}{2}}^2\exp\ke{-\frac\beta2 E_+ + 2i\lambda t_- E_-^2 + it_+ E_-  -\frac2\Gamma |E_-|}
\end{align}
based on the weakly coupled limit $|\lambda|\ll 1/E_\beta$.
For large $D$, the energy band $\Lambda=E_{\max}-E_{\min}$ is much larger than the bandwidth $\Gamma$. So we can calculate the integral in the approximation of flat spectrum difference as
\begin{align}
\sum_{a\neq b}\approx
D^2\int_0^\Lambda dE_a dE_b\,\rho(E_a)\rho(E_b)
\approx D^2\int_0^{2\Lambda} dE_+\,\rho(E_+/2)\int_{-\infty}^{+\infty}dE_-.
\end{align}
Then, the off-diagonal part is simplified as
\begin{align}
&D\int_0^{2\Lambda}dE_+\,A\kc{\frac{E_+}{2}}^2\rho\kc{\frac{E_+}{2}} \exp\ke{-\frac\beta2 E_+} \int_{-\infty}^{+\infty}dE_- \exp\ke{2i\lambda t_- E_-^2 +  it_+E_- -\frac2\Gamma |E_-|} \nn\\
=&2\kd{D\int_0^{\Lambda}dE\, A(E)^2\rho(E) e^{-\beta  E}} \sqrt{\frac{\pi }{-8i\lambda t_-}} 
\kd{ g\left(\frac{2/\Gamma-it_+}{\sqrt{-8i\lambda t_-}}\right) +g\left(\frac{2/\Gamma+it_+}{\sqrt{-8i\lambda t_-}}\right)},
\end{align}
where
\begin{align}
g(z)=e^{z^2}\text{erfc}(z)=e^{z^2}\kc{1-\frac{2}{\sqrt{\pi }}\int _0^z dx e^{-x^2}}.
\end{align}
Let $A^2=Z[\beta]^{-1} D\int_0^{\Lambda}dE\, A(E)^2\rho(E) e^{-\beta  E}$. The retarded correlator is
\begin{align}
G_{LR}^R(t_1,t_2)\approx 4 A^2 \Im\ke{  \sqrt{\frac{\pi }{-i8\lambda t_-}} \kd{ g\left(\frac{2/\Gamma-it_+}{\sqrt{-8i\lambda t_-}}\right) + g\left(\frac{2/\Gamma+it_+}{\sqrt{-8i\lambda t_-}}\right)}}.
\end{align}
Asymptotically,
\begin{align}
G_{LR}^R(t,-t)\to 2A^2
\begin{cases}
32 \Gamma ^3 \lambda  t_-\frac{4-3 \Gamma ^2 t_+^2}{(4+\Gamma ^2 t_+^2){}^3}, & t_-\to0\\
\sqrt{\frac{\pi}{\lambda t_-}}, & t_-\to \infty
\end{cases}.
\end{align}
whose exponents are the same as the conformal result in Fig.~\ref{fig:GLRCA}.

By replacing $it_-\to it_--\frac\beta4$, we obtain the retarded correlator of the $\TT$ deformed TFD state as
\begin{align}
G_{LR}^R(t_1,t_2)_\lambda \approx 4 A^2 \Im\ke{  \sqrt{\frac{\pi }{2\lambda (\beta-4it_-)}} \kd{ g\left(\frac{2/\Gamma-it_+}{\sqrt{2\lambda (\beta-4it_-)}}\right) + g\left(\frac{2/\Gamma+it_+}{\sqrt{2\lambda (\beta-4it_-)}}\right)}}.
\end{align}

\section{$\TT$ deformation on Schwarzian theory}\label{Sec:Schwarzian}

In this section, we consider the $\TT$ deformation on an eternal black hole in Jackiw-Teitelboim (JT) gravity as \cite{Maldacena:2016upp,Engelsoy:2016xyb}
\begin{align}
I=\frac1{16\pi G}\kd{\int d^2x\sqrt{-g}\Phi(R+2)+2\int_b dx \sqrt{-h}\Phi_b (K-1)}
\end{align}
with the boundary condition
\begin{align}
ds^2_h=-dt^2/\epsilon^2,\quad \Phi_b=\Phi_r/\epsilon,
\end{align}
where $b$ denotes the boundary, $h$ is the induced metric on the boundary, $\Phi_b$ is the value of the dilaton $\Phi$ on the boundary, and $\epsilon$ is the UV cutoff.
We have introduced a counterterm to cancel the divergence in the exterior curvature $K$.
By integrating out the dilaton $\Phi$, we have $R+2=0$. The solution is an AdS$_2$ space. In the global coordinate and the Rindler coordinate, the metric reads
\begin{align}
ds^2=\frac{-d\nu^2+d\sigma^2}{\sin^2\sigma}=-\sinh^2\rho d\varphi^2+d\rho^2.
\end{align}
We consider two boundaries $L$ and $R$ with reparametrization $\kc{\varphi_L(t),\rho_L(t)}$ and $\kc{\varphi_R(t),\rho_R(t)}$ respectively. To satisfy the boundary condition, the reparametrizations are expanded as
\begin{align}\label{FiniteCutOffL}
\sinh\rho_L(t)=&-\frac1{\epsilon\varphi_L'(t)}- \frac{\epsilon\varphi_L''(t)^2}{2\varphi_L'(t)^3} + \mathcal O(\epsilon^2),\\
\label{FiniteCutOffR}
\sinh\rho_R(t)=&\frac1{\epsilon\varphi_R'(t)}+ \frac{\epsilon\varphi_R''(t)^2}{2\varphi_R'(t)^3}  + \mathcal O(\epsilon^2).
\end{align}
The action is then reduced to the two-sited Schwarzian theory:
\begin{align}
I
=&\frac1{8\pi G}\int_{L,R} dx \sqrt{-h}\Phi_b (K-1) \label{SurfaceTerm}\\
=&-C\int dt\kd{\Sch\kc{-\coth\frac{\varphi_L}{2},u}+\Sch\kc{\tanh\frac{\varphi_R}{2},u}}+\mathcal O[\epsilon^2]\\
=&\frac C2\int dt\kd{\frac{\varphi_L''(t){}^2}{\varphi_L'(t){}^2}+ \varphi_L'(t){}^2+\frac{\varphi_R''(t){}^2}{\varphi_R'(t){}^2}+ \varphi_R'(t){}^2}+\text{surface term},\quad C=\frac{\Phi_r}{8\pi G}.
\end{align}
Similar to the argument presented in \cite{Maldacena:2018lmt}, the action has SL(2) gauge symmetry, and the gauge charges vanish. Therefore, the solution can be transformed into the $LR$-symmetric form $\varphi_L(u)=\varphi_R(t)=\varphi(t)$. Following \cite{Gross:2019ach}, we will consider the $\TT$ deformation and use Ostrogradsky formalism to write down the canonical variables \cite{Woodard:2015zca}
\begin{equation}\label{Canonical}\begin{split}
& q_{1}=\varphi,\quad
q_{1}=\varphi',\\
& p_{1}=\frac{\partial L}{\partial \varphi'}-\partial_t \frac{\partial L}{\partial \varphi''}
=C\kd{\frac{\varphi''^2}{\varphi'^3}+\varphi'-\frac{\varphi^{(3)}}{\varphi'^2}}
,\quad
p_{2}=\frac{\partial L}{\partial \varphi''}=C\frac{\varphi''}{\varphi'^2}.
\end{split}\end{equation}
The Hamiltonian in Ostrogradsky formalism is
\begin{align}
H_0=H_L+H_R=p_{1} q_{2} +\frac{1}{4C} p_{2}^2 q_{2}^2-C q_{2}^2.
\end{align}
The solution of the TFD state at inverse temperature $\beta$ is
\begin{align}\label{ParameterizationTFD}
\varphi(t)=\varphi_{\beta}(t)=\frac{2\pi  t}{\beta },
\end{align}
and all the canonical variables are determined by (\ref{Canonical}). With (\ref{FiniteCutOffL}) and (\ref{FiniteCutOffR}), the solution of the reparametrization describes two boundary trajectories on the constant radius in the Rindler patch of AdS$_2$ space, which corresponds to a wormhole connecting the two boundaries from a higher-dimensional perspective.

We first consider a general deformation:
\begin{align}
H_\lambda=f(H_0).
\end{align}
Under the deformation, the canonical relations are determined by the deformed Hamiltonian equation as
\begin{align}
q'_{i}=\frac{\partial H_\lambda}{\partial p'_{i}},\quad
p'_{i}=-\frac{\partial H_\lambda}{\partial q'_{i}}.
\end{align}
Given a solution of the Hamiltonian equation of $H_0$, such as the $\varphi_\beta(t)$ in (\ref{ParameterizationTFD}), we can find a solution of $H_\lambda$ from (\ref{DeformedSolution}). We introduce the dynamical time as
\begin{align}
T=kt, \quad k=f'(H_0[\varphi_\beta(t)]),
\end{align}
where $H_0[\varphi_\beta(t)]=4\pi^2 C/\beta^2$ refers to the value of $H_0$ on the canonical variables determined by the solution in (\ref{ParameterizationTFD}) with the canonical relation in (\ref{Canonical}). A solution of the Hamiltonian equation of $H_\lambda$ is
\begin{align}\label{DeformedSchSolution}
&\varphi(t)=\varphi_\beta(kt)=\varphi_{\beta/k}(t)=2\pi kt/\beta,\\
&q_{1}=2\pi kt/\beta,\quad
q_{2}=2\pi/\beta,\quad
p_{1}=2\pi C/\beta,\quad
p_{2}=0,
\end{align}
and the energy is $H_\lambda[\varphi_\beta(kt)]=f(4\pi^2 C/\beta^2)$.

We select the $\TT$ deformation as
\begin{align}
H_\lambda=f(H_0)=\frac{1-\sqrt{1-8 H_0 \lambda }}{4 \lambda }.
\end{align}
Then, $k=1/\sqrt{1-32\pi^2 \lambda C/\beta^2}$ and $ H_\lambda[\varphi_\beta(kt)]=\kc{1-\sqrt{1-32 \pi ^2\lambda C/\beta ^2}}/4 \lambda$. So, the weakly coupled limit in Sec.~\ref{Sec:QM} means $|\lambda|\ll \beta^2/C$ and $k\approx 1$ here.

The solution (\ref{DeformedSchSolution}) has two interpretations that correspond to the two strategies separately in Sec.~\ref{Sec:QM}. First, recall that the local state $\rho(t)$ in (\ref{QuenchedLocal}) is a thermal state of the undeformed Hamiltonian $H_0$. Thus, the solution $\varphi(t)=\varphi_{\beta/k}(t)$ is interpreted as the $\TT$ quenched TFD state $\ket{\Psi(t)}$ at the inverse temperature $\beta/k$ in (\ref{QuenchedTFD}). Second, recall that the local state $\rho_\lambda(t)$ in (\ref{DeformedLocal}) is a thermal state of the deformed Hamiltonian $f(H_0)$, in which the dynamical time is $kt$ as well. Thus, the solution $\varphi(t)=\varphi_\beta(kt)$ is interpreted as the $\TT$ deformed TFD state $\ket{\Psi_\lambda(t)}$ at the inverse temperature $\beta$ in (\ref{DeformedTFD}).

\begin{figure}
	\centering
	\includegraphics[height=0.25\linewidth]{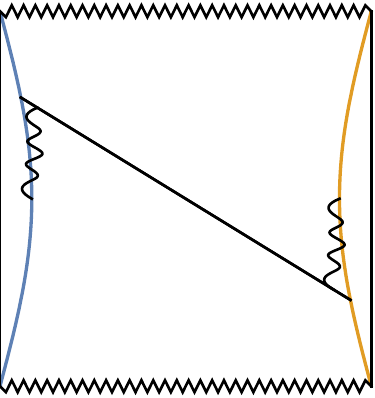}
	\caption{Witten diagram in retarded correlator $G^R_{LR}$ at first the order of $\lambda$ in the global coordinate of AdS$_2$. The solid line represents the propagator of matter field and the wavy line represents the propagator of boundary graviton (reparametrization model). The boundaries are spacelike separated.}
	\label{fig:Bulk}
\end{figure}

The deformed solution (\ref{DeformedSchSolution}) is related to the undeformed solution (\ref{ParameterizationTFD}) by rescaling the time. By plugging (\ref{DeformedSchSolution}) into (\ref{FiniteCutOffL}) and (\ref{FiniteCutOffR}) respectively, we know that the $\TT$ deformation moves the two boundaries into the bulk but keeps them spacelike separated, which agrees with the fact that the deformation is irrelevant. Thus, the causal correlation found in Sec.~\ref{Sec:QM} is not associated with the causal structure of a semiclassical wormhole and is instead similar to the ``quantum traversable wormhole''
\footnote{Notice that the traversability of the semiclassical traversable wormholes in Ref.~\cite{Gao:2016bin} is different from the traversability of the ``quantum traversable wormhole'' in Ref.~\cite{Gao:2018yzk}. The former happens after the scrambling time and is related to the spacetime structure of the bulk. The latter occurs in the interference regime much later than the scrambling time. It is generated by the superposition of the bulk states and is not related to the spacetime structure.}
in Ref.~\cite{Gao:2018yzk}.
{Without the $\TT$ deformation, the vanishing of the retarded correlator is originated from the perfect cancellation between the two propagators $\avg{O_L^{(0)}O_R^{(0)}}$ and $\avg{O_R^{(0)}O_L^{(0)}}$, which are dual to the process of a virtual particle traveling from $R$ to $L$ and from $L$ to $R$ in bulk respectively.} With the $\TT$ deformation, the virtual particle can release two gravitons that annihilate on the boundaries via the $H_LH_R$ term in the $\TT$ deformation, as shown in Fig.~\ref{fig:Bulk}. The propagators acquire different factors, resulting in the propagation of real particles.

More precisely, we can directly calculate retarded correlator $G^R_{LR}(t_1,t_2)$ at the first order of $\lambda$ by using the Schwarzian action. Taking the $G^R_{LR}(t_1,t_2)$ in (\ref{GLRInstantPerturb}) as an example, where the $\TT$ quench is applied instantaneously, we consider the reparametrization mode $\varphi(t)=t+\varepsilon(t)$ and expand the dynamical part of the Euclidean Schwarzian action, the correlator, and the nonlocal $\TT$ term with respect to $\varepsilon(t)$ as
\begin{align}
I_\varepsilon=&\frac{C}{2}\int d\tau \kc{\varepsilon''(\tau)^2-\varepsilon'(\tau)^2+\mathcal O[\varepsilon^3]},\\
\avg{O(\tau_1)O(\tau_2)}=& \kd{\frac{\left(1+\varepsilon '(\tau_1)\right) \left(1+\varepsilon'(\tau_2)\right)}{4} \csc^2 \frac{\tau _1-\tau _2+\varepsilon (\tau_1)-\varepsilon (\tau_2)}{2}}^{\Delta },\\
-4\lambda H_LH_R=&\mathcal O[\varepsilon]-4\lambda C^2 \left(\varepsilon^{(3)}(0)+\varepsilon'(0)\right) \left(\varepsilon^{(3)}(\pi)+\varepsilon'(\pi)\right)+\mathcal O[\varepsilon^3],\quad \beta=2\pi.
\end{align}
The quadratic term in $I_\varepsilon$ gives the propagator of reparametrization model $\avg{\varepsilon(\tau)\varepsilon}=-(\pi-|\tau|) (\pi-|\tau| +2 \sin|\tau| )/(4\pi C)$ \cite{Maldacena:2016upp}. The brackets $\avg{\varepsilon\varepsilon\varepsilon\varepsilon}$ in the commutator $-4\lambda \langle [[H_LH_R,O_L(t_1)], O(t_2)]\rangle$ are factorized into $\avg{\varepsilon\varepsilon}\avg{\varepsilon\varepsilon},$ as shown in Fig.~\ref{fig:Bulk}. The $\mathcal O[\varepsilon^3]$ term in $I_\varepsilon$ and the $\mathcal O[\varepsilon]$ term in $-4\lambda H_LH_R$ do not contribute to the commutator. The final result is (\ref{GLRInstantResult}) at the tree level.

By Legendre transforming the deformed Hamiltonian and letting $q_1=\varphi,\ q_2=e^{\phi}$, we obtain the deformed Lagrangian as
\begin{align}
\L_\lambda=\frac{C e^{\phi } \left(\varphi '^2+\phi '^2\right)}{\varphi '}+\frac{\left(e^{\phi }-\varphi '\right)^2}{8 \lambda  \varphi 'e^{\phi } }.
\end{align}
Solving $\varphi$ and substituting it in the Lagrangian, we have
\begin{align}\label{DeformedLagrangian}
\L_\lambda=\frac{\sqrt{\left(1+8 C \lambda  \phi '^2\right)\left(1+8 C \lambda  e^{2 \phi }\right) }-1}{4\lambda},
\end{align}
which agrees with the $\TT$ deformation of the Liouville QM $\L=\frac C2(\phi_L'^2+\phi_L'^2+e^{2\phi_L}+e^{2\phi_R})$ when $\phi_L=\phi_R=\phi$, as given in (\ref{TTLag}).

To keep the correction of finite cutoff $\epsilon$, we substitute (\ref{FiniteCutOffL}) and (\ref{FiniteCutOffR}) into the action (\ref{SurfaceTerm}) and obtain the Lagrangian as
\begin{align}
\L_\epsilon=C\sum_{s=L,R}\frac12\left(\varphi _s'{}^2+\frac{\varphi _s''{}^2}{2 \varphi _s'{}^2}\right)+\frac{\epsilon ^2}{8} \left(-\frac{57 \varphi _s''{}^4}{\varphi _s'{}^4}-\varphi _s'{}^4+\frac{8 \varphi _s'''{}^2}{\varphi _s'{}^2}-6 \varphi _s''{}^2\right)+\mathcal O\left[\epsilon ^3\right].
\end{align}
Substituting the solution (\ref{ParameterizationTFD}) in the above Lagrangian, we get $\L_\epsilon=4 \pi ^2 C\beta ^{-2}-16 \pi ^4 C^2 \lambda/\beta^{-4}$, where we use $\lambda=2\pi \epsilon^2 G/\Phi_R$, based on \cite{Hartman:2018tkw,Gross:2019uxi}. However, if we substitute the deformed solution (\ref{DeformedSchSolution}) into (\ref{DeformedLagrangian}), we get $\L_\lambda=4 \pi ^2 C\beta ^{-2}-32 \pi ^4 C^2 \lambda/\beta^{-4}$, which is different form $\L_\epsilon$. Therefore, the effect of the nonlocal $\TT$ deformation considered in this study is not simply equivalent to moving the boundaries into the bulk \cite{McGough:2016lol}. Rather, it couples the two boundaries and leads to nonlocal dynamics.

\section{Summary and prospect}\label{Sec:Summary}

In this study, we reformulate the $\TT$ deformation of multiple systems in the (0+1) dimension in terms of the dynamical coordinate transformation, which originated from 2D $\TT$ deformation of quantum field theories \cite{Dubovsky:2018bmo,Dubovsky:2017cnj, Cardy:2018sdv,Conti:2018tca}. In 2D $\TT$ deformation, the deformed quantum field theory is equivalent to the seed theory coupling with 2D massive gravity. We generalize the philosophy to a (0+1) dimension quantum mechanic system. By using the known fact of $\TT$ deformed SYK model \cite{Gross:2019uxi}, we obtain the so-called 1D massive gravity formalism and then obtain the Hamiltonian for $\TT$ deformation of multiple systems in the (0+1) dimension. It has been confirmed that it is equivalent to the $\TT$ deformation by flow equation in 2D deformed quantum field theory. Given a solution of the original theory, we can find an explanation of the deformed theory related to the original resolution by time rescaling. Motivated by this rescaling, we introduce the dynamical tetrad acting as one-dimensional gravity. By integrating the dynamical tetrad, we can obtain the deformed action. The $\TT$ deformation of multiscalar theory follows a similar form.

The $\TT$ deformation of bi-systems effectively couples the local systems. We further consider the TFD states on the bi-system. The signals injected into one system at a particular time can appear from the other at the reversal time. The time of best traversal scales is $\beta^2/\lambda$ in conformal QM. In the SYK model, our analytical result at the large-$q$ limit was close to the result obtained from exact diagonalization. For the theory satisfying ETH, we find that the traversal is dependent mainly on the bandwidth of the operator carrying the signal.

Finally, we study such $\TT$ deformation on two-sited Schwarzian action, which describes the leading nonconformal dynamics of the eternal black hole in JT gravity. We obtain the deformed Lagrangian and find that the deformed solution is an external black hole with rescaled time, whose two boundaries are spacelike separated. It shows that the regenesis found in the bi-QM system is not associated with the causal structure of a semiclassical wormhole.

In this study, we focus on the regenesis phenomenon of the TFD state under $\TT$ deformation. Because $\TT$ coupling is directly related to energy, the energy transport also merits investigation in the future \cite{Xian:2019qmt}. Our study of the regenesis phenomena under the $\TT$ deformation gives a new perspective of the information process, and the causal structure of $\TT$ deformed field theories. We expect that the regenesis phenomena under $\TT$ deformation are common in highly entangled states because this deformation is not required to match the entanglement structure of the TFD state. It is natural to extend the $\TT$ deformation to the CFT$_2$ with multiple fields and check the regenesis of the deformed TFD states \cite{He:2019vzf,He:2020udl,He:2020qcs}. In terms of \cite{Bzowski:2020umc}, one can choose proper two-sided $\TT$ coupling to reconstruct the bulk geometry of the deformed TFD state and compare the correlators from gravity and field theory.

\section*{Acknowledgments}
We thank Avik Chakraborty, Chris Lau, Yuan Sun, Tadashi Takayanagi, Hao Ouyang, and Long Zhao for valuable discussions related to this work. We particularly appreciate the conversation with Hao Ouyang about how the $T\bar{T}$ can be regarded as a dynamical change of coordinates. S.H. also appreciate the financial support from Jilin University and Max Planck Partner group, as well as the National Natural Science Foundation of China under Grants No.~12075101 and No.~12047569. Z. Y. X. acknowledges support from the National Natural Science Foundation of China under Grants No.~11875053 and No.~12075298 and the Deutsche Forschungsgemeinschaft (DFG, German Research Foundation) under Germany's Excellence Strategy through the W\"urzburg-Dresden Cluster of Excellence on Complexity and Topology in Quantum Matter ct. qmat (EXC 2147, project id 390858490).

\end{document}